# Analysis of Entanglement Measures and LOCC Maximized Quantum Fisher Information of General Two Qubit Systems

*Volkan Erol[1,2], Fatih Ozaydin[3] and Azmi Ali Altintas[4]

1 Institute of Science, Okan University, Istanbul, Turkey

2 Provus A MasterCard Company R&D Center, Istanbul, Turkey

3 Department of Information Technologies, Isik University, Istanbul, Turkey

4 Department of Electrical Engineering, Okan University, Istanbul, Turkey

*: main correspondence can be made with Volkan Erol volkan.erol@gmail.com

**Abstract**

**Entanglement has been studied extensively for unveiling the mysteries of non-classical correlations between quantum systems. In the bipartite case, there are well known measures for quantifying entanglement such as concurrence, relative entropy of entanglement *(REE)* and negativity, which cannot be increased via local operations. It was found that for sets of non-maximally entangled states of two qubits, comparing these entanglement measures may lead to different entanglement orderings of the states. On the other hand, although it is not an entanglement measure and not monotonic under local operations, due to its ability of detecting multipartite entanglement, quantum Fisher information (QFI) has recently received an intense attraction generally with entanglement in the focus. In this work, we revisit the state ordering problem of general two qubit states. Generating a thousand random quantum states and performing an optimization based on local general rotations of each qubit, we calculate the maximal QFI for each state. We analyze the maximized QFI in comparison with *concurrence, REE* and *negativity* and obtain new state orderings. We show that there are pairs of states having equal maximized QFI but different values for *concurrence, REE* and *negativity* and vice versa.**

## Introduction

Quantification and manipulation of entanglement between two or more parties have been one of the central research directions in quantum mechanics since the seminal EPR paper [1] and especially with the emergence of quantum information theory. Maximum entanglement of two qubits shared between two parties can appear as one of four orthogonal states, called Bell states, which can be transformed into each other by local operations and classical communication (LOCC) [2] while such a transformation becomes impossible in multipartite setting, where states fall into inequivalent classes such as GHZ [3], Cluster [4], W [5], and Dicke [6] states. Although efficient generation of GHZ and Cluster states of arbitrary sizes has been achieved so far [7,8] currently there is an intense effort for W [9-15] and Dicke states [16]. What is more, a general measure for quantifying multipartite entanglement is yet to be found. A basic criterion for a measure to be a valid entanglement measure is the monotonicity, i.e. since entanglement cannot be increased via LOCC, the value that the measure provides should not be increased via LOCC. In the bipartite case, there are well-known entanglement measures, such as concurrence, relative entropy of entanglement (REE) and negativity [17-22]. Concurrence and negativity measures are based on the eigenvalues of the density matrix (after some transformations for Concurrence). REE is based on the distance of the state to the closest separable state. Although the value of such measures for separable states turns to be 0, and 1 for maximally entangled states, it was found that this is not the case for the states in between [18,23-26]: There are pairs of states that have equal values of one measure but different values of another measure; and even more surprisingly, for one measure, the value of the first state is larger than the second but for another measure, the value of the first state is smaller than the second. This result on ordering the states with respect to different entanglement measures can be interpreted as, each of these measures is probably reflecting some other peculiarity of entanglement.

When it comes to detect entanglement, quantum Fisher information (QFI) has been found to be a useful tool [27]. In particular, if a state exceeds the shot-noise level (SNL) that the best separable states can achieve, which is 1, then that state is entangled. For an N particle state, the fundamental limit or the Heisenberg limit (HL) of QFI is N, which is achievable by GHZ states, for example. Therefore, besides providing information for the phase sensitivity of the state with respect to SU(2) rotations in the context of quantum metrology [28-38]; QFI has received an intense attention with the entanglement in the focus [39-45,52]. There are attempts to find a QFI based general entanglement measure that can quantify not only bipartite but also multipartite states and even bound entangled states [43]. A recent work has shown that bound entangled states can reach the Heisenberg level [44]. Since QFI provides information for the sensitivity of a state with respect to changes in the state, including unitary operations, it is naturally non-invariant with respect to changes in general. Therefore when comparing the QFI of a state with entanglement measures (which are invariant with respect to unitary rotations), it would be plausible to perform an optimization to find the maximal QFI of the state over all possible local unitary rotations.

Since QFI plays an important role in various aspects of entanglement, in this work, we extend the state ordering problem which was studied in the context of the standard entanglement measures, with the maximized QFI. We generate a thousand two-qubit states, calculate their concurrence and negativity values, run a simulation for obtaining the relative entropy of entanglement values, calculate QFI and maximize the QFI with respect to general local unitary operations and finally analyze these results. We find that there are pairs of states such that maximized QFI of the first state can be larger than the second, whereas any standard entanglement measure of the second being larger than the first, and vice versa. We believe that our work may contribute to this lively field.

**Results**

We generated a thousand random two-qubit states of which 625 turns to be separable and the rest to be entangled, in accordance with the results of [18, 46]. Calculating the entanglement measures and maximized QFI (MQFI) values for each state, we found pairs of states $\{\rho_1,\rho_2\}$, in each possible class of different orderings with respect to standard entanglement measures and MQFI, as presented in Table 1. Since any entanglement measure of the separable 625 states turns to be zero, we treat them to be in the same set, i.e. *Entanglement($\rho_1$)=Entanglement($\rho_2$)=0*. For the entangled states which possibly have different orderings with respect to concurrence, REE and entanglement (as presented in [18,23-26]), in Table 1 we can treat pairs of entangled states in one of three possible classes.

In Figure 1, we present the results of state orderings with respect to (a) concurrence, (b) negativity and (c) REE versus (blue dots) QFI, (red dots) LOCC Maximized QFI and (green dots) LOCC Minimized QFI. A main result presented in this figure is that the results of [18,23-26] can be reflected the best when QFI is maximized, especially when considering the ordering with respect to REE. This result would suggest that when studied in the context of entanglement, QFI should be LOCC maximized, because the boundaries of the region of not the blue or the green dots but only of the red dots reflect the comparison results, presented in Figure 1 of Ref.[25]. In particular, only the red dots are compatible with the curves of comparing REE with concurrence and comparing REE with negativity.

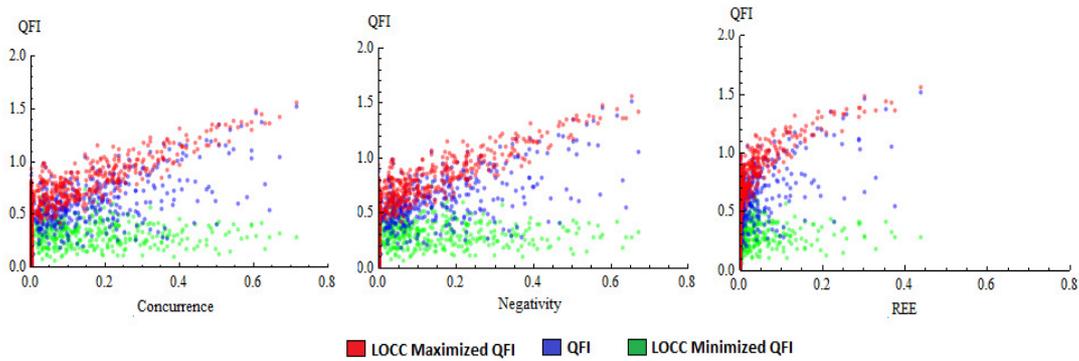

**Figure 1 |** Comparison of (red) maximized QFI, (blue) QFI and minimized (green) QFI with respect to entanglement measures: (a) Concurrence, (b) Negativitiy and (c) relative entropy of entanglement of one thousand random states.

**Discussion**

It was found that in two-qubit systems, state ordering with respective to quantification of entanglement, highly depends on the entanglement measure used [18,23-26]. For example, considering a pair of states, negativity of the first state can be larger than the negativity of the second state, whereas the concurrence of the second state can be larger than the concurrence of the first state. This finding opened new insights in our understanding the entanglement in the bipartite case and has the potential to give rise to even more interesting results in the multipartite settings. Regarding the detection of multipartite entanglement and providing information for the phase sensitivity of a state with respect to SU(2) rotations, quantum Fisher information (QFI) is a practical tool and is being studied within the context of entanglement [27,43].

Incorporating QFI, we have extended the state ordering problem of general two qubit systems. Since QFI is not monotonic under LOCC, we have maximized (and minimized) QFI with respect to general Euler rotations of each qubit. We have chosen the maximal (and minimal) values of QFI and compared these values with the values obtained by the entanglement measures for each state. The main result is that regardless of being maximized or minimized by LOCC, QFI of two-qubit states behave similar to the entanglement measures in such a comparative analysis: For a pair of states, QFI of the first state can be larger than the QFI of the second whereas any entanglement measure of the second state can be larger than the same measure of the first.

On the other hand, effect of maximizing QFI becomes significant for revealing the dynamics of comparison of REE with the other two measures, in the comparison of QFI with these measures. An interesting result is that QFI of almost all of the generated random states were both maximized and minimized under LOCC via general Euler rotations, even with large steps (each being $\frac{\pi}{2}$) when scanning the whole space, $[0, 2\pi]$. The few states that remained unchanged were both maximized and minimized as the step size was chosen to be $\frac{\pi}{3}$.

Observing that QFI of all the random states are both maximized and minimized via local SU(2) operations might open new questions for finding non-maximally mixed but QFI non-invariant states under rotations in a similar vein to quantum discord (QD), since QD of SU(2) invariant states has been recently studied [48,49].

We believe that our results can be useful for understanding the relation between QFI and entanglement not only in the bipartite but also in the multipartite settings.

**Methods**

**Concurrence**

For a density matrix $\rho$, concurrence is given in [17] as

$$C(\rho) = \max\{0, \lambda_1 - \lambda_2 - \lambda_3 - \lambda_4\} \quad (1)$$

where $\lambda_i$'s are the square roots of the eigenvalues of the matrix

$$\rho(\sigma_y \otimes \sigma_y)\rho^*(\sigma_y \otimes \sigma_y) \quad (2)$$

in decreasing order and $\sigma_i$'s are the Pauli spin operators and $\rho^*$ is the complex conjugate of $\rho$. $C(\rho)$ ranges between 0 for a separable state and 1 for a maximally entangled state.

**Negativity**

Negativity can be considered a quantitative version of the Peres-Horodecki criterion [50,51]. The negativity for a two-qubit state σ is defined as [18,19,43]:

$$N(\rho) = 2\sum_i \max(0, -\mu_i) \quad (3)$$

where $\mu_i$ is the negative eigenvalues of the partial transpose of $\rho$. Negativity also ranges between 0 for a separable state and 1 for a maximally entangled state. As shown by Vidal and Werner [19], the negativity is an entanglement monotone which means that it can be considered as a useful measure of entanglement.

**Relative Entropy of Entanglement (REE)**

Relative Entropy of Entanglement (REE) of a given state σ, which is defined by Vedral et al [21,22] as the minimum of the quantum relative entropy

$S(\rho||\sigma) = Tr(\rho log\rho - \rho log\sigma)$ taken over the set D of all separable states σ, namely

$$E(\rho) = \min_{\rho \in D} S(\rho \parallel \sigma) = S(\rho \parallel \bar{\sigma}) \tag{4}$$

where $\bar{\sigma}$ denotes a separable state closest to ρ. In general, REE is calculated numerically using the methods described in [22,46].

**Quantum Fisher Information**

Defining the fictitious angular momentum operators on each qubit in each direction,

$$J_{\vec{n}} = \sum_{\alpha=x,y,z} \frac{1}{2} n_\alpha \sigma_\alpha \tag{5}$$

and the quantum Fisher information of a state $\rho$ of N particles with eigenvalues $p_i$ and the associated eigenvectors $|i\rangle$ in each direction as $F(\rho, J_{\vec{n}})$, the maximal mean quantum Fisher information of the state can be found as

$$\bar{F}_{max} = \frac{1}{N} \max_{\vec{n}} F(\rho, J_{\vec{n}}) = \frac{\lambda_{max}}{N}. \tag{6}$$

where $\lambda_{max}$ is the largest eigenvalue of the 3x3 symmetric matrix C, of which elements can be calculated by [45]

$$C_{kl} = \sum_{i \neq j} \frac{(p_i - p_j)^2}{p_i + p_j} \left[ \langle i | J_k | j \rangle \langle j | J_l | i \rangle + \langle i | J_l | j \rangle \langle j | J_k | i \rangle \right], \tag{7}$$

where $k,l \in \{x,y,z\}$.

**Random State Density Matrix Generation**

In this work, the density matrices of the random states are generated as follows:

$$\rho = VPV^\dagger \tag{8}$$

Where $P = diag(\lambda_i)$ diagonal matrix of eigenvalues and V is the unitary matrix [18]. To get such random state we have used the methods in the QI package [50].

**Optimization of QFI**

We have performed the maximization via general rotations of each qubit in the Euler representation

$$U_{Rot}(\alpha,\beta,\gamma) = U_x(\alpha)U_z(\beta)U_x(\gamma) \quad (9)$$

where the rotations about axes are defined as $U_j(\alpha) = \exp(-i\alpha\frac{\sigma_j}{2})$, $j \in \{x,z\}$, with arbitrary three angles for each qubit between $[0, 2\pi]$, each with steps of $\theta$ degrees, resulting $O(\left(\frac{2\pi}{\theta}\right)^6)$ QFI calculations. We have found that choosing the steps as $\theta = \frac{\pi}{2}$ is sufficient for a good optimization such that the picture, whereas narrowing the steps could possibly result a better optimization, with the cost of an increase in the running time of the simulation. For each state, we also found the minimal QFI in the same general rotation process which is shown as green in Figure 1.

According to the results that we obtained, for the chosen a thousand random states, %98 of the QFI values are maximized and %99 of the QFI values are minimized with $\theta = \frac{\pi}{2}$ and the all the rest with $\theta = \frac{\pi}{3}$.

**References**

[1] A. Einstein, B. Podolsky and N. Rosen, Can quantum-mechanical description of physical reality be considered complete?, *Phys. Rev.* 47 (1935) 777–780.

[2] M. A. Nielsen and I.L. Chuang, *Quantum Computation and Quantum Information,* Cambridge University Press (Cambridge) (2000).

[3] D. M. Greenberger, M.A. Horne, and A. Zeilinger, *in Bell's Theorem, Quantum Theory, and Conceptions of the Universe, edited by M. Kafatos* (Kluwer Academics, Dordrecht, The Netherlands 1989), p. 73.

[4] R. Raussendorf and H.J. Briegel, A One-Way Quantum Computer, Phys. Rev. Lett. 86, 5188 (2001).

[5] W. Dür, Multipartite entanglement that is robust against disposal of particles, *Phys. Rev. A* 63, 020303 (2001).

[6] R. H. Dicke, Coherence in Spontaneous Radiation Processes, *Phys. Rev.* 93, 99 (1954).

[7] D. E. Browne and T. Rudolph, Resource-Efficient Linear Optical Quantum Computation, *Phys. Rev. Lett.* 95, 010501 (2005).

[8] A. Zeilinger, M. A. Horne, H. Weinfurter, and M. Zukowski, Three-Particle Entanglements from Two Entangled Pairs, *Phys. Rev. Lett.* 78, 3031 (1997).

[9] T. Tashima, S. K. Ozdemir, T. Yamamoto, M. Koashi, and N. Imoto, Elementary optical gate for expanding an entanglement web, *Phys. Rev. A 77,* 030302 (2008).

[10] T. Tashima, S. K. Ozdemir, T. Yamamoto, M. Koashi, and N. Imoto, Local expansion of photonic W state using a polarization-dependent beamsplitter, *New J. Phys. A* 11, 023024 (2009).

[11] T. Tashima, T. Wakatsuki, S. K. Ozdemir, T. Yamamoto, M. Koashi, and N. Imoto, Local Transformation of Two Einstein-Podolsky-Rosen Photon Pairs into a Three-Photon W State, *Phys. Rev. Lett.* 102, 130502 (2009).

[12] S. K. Özdemir, E. Matsunaga, T. Tashima, T. Yamamoto, M. Koashi, and N. Imoto, An optical fusion gate for W-states, *New J. Phys.* 13, 103003 (2011).

[13] S. Bugu, C. Yesilyurt and F. Ozaydin, Enhancing the W-state quantum-network-fusion process with a single Fredkin gate, *Phys. Rev. A* 87, 032331 (2013).

[14] C. Yesilyurt, S. Bugu and F. Ozaydin, An Optical Gate for Simultaneous Fusion of Four Photonic W or Bell States, *Quant. Info. Proc.* 12, 2965 (2013).

[15] F. Ozaydin, S. Bugu, C. Yesilyurt, A. A. Altintas, M. Tame, Ş. K. Özdemir, Fusing multiple W states simultaneously with a Fredkin gate, *Phys. Rev. A* 89, 042311 (2014).

[16] T. Kobayashi, R. Ikuta, Ş. K. Özdemir, M. Tame, T. Yamamoto, M. Koashi and N. Imoto, Universal gates for transforming multipartite entangled Dicke states, *New J. Phys.* 16 023005 (2014).

[17] W. K. Wootters, Entanglement of Formation of an Arbitrary State of Two Qubits, *Phys. Rev. Lett.* 80, 2245 (1998).

[18] J. Eisert and M. B. Plenio, A comparison of entanglement measures, *J. Mod. Opt*. 46, 145-154 (1999).

[19] G. Vidal and R. F. Werner, Computable measure of entanglement, *Phys. Rev. A* 65, 032314 (2002).

[20] K. Zyczkowski, P. Horodecki, A. Sanpera, and M. Lewenstein, Volume of the set of separable states, *Phys. Rev. A* 58, 883 (1998).

[21] V. Vedral, M. B. Plenio, M. A. Rippin and P. L. Knight, Quantifying Entanglement, *Phys.Rev. Lett.* 78 2275 (1997).

[22] V. Vedral and M. B. Plenio, Entanglement measures and purification procedures, *Phys. Rev. A* 57 1619 (1998).

[23] F. Verstraete, K. M. R. Audenaert, J. Dehaene, and B.De Moor, A comparison of the entanglement measures negativity and concurrence, *J. Phys. A* 34, 10327 (2001).

[24] A. Miranowicz and A. Grudka, Ordering two-qubit states with concurrence and negativity, *Phys. Rev. A* 70, 032326 (2004).

[25] A. Miranowicz and A. Grudka, A comparative study of relative entropy of entanglement, concurrence and negativity, *J. Opt. B: Quantum Semiclass. Opt.* 6, 542–548 (2004).

[26] A. Miranowicz, e-prints quant-ph/0402023 and quant-ph/0402025 (2004).

[27] L. Pezze, and A. Smerzi, Entanglement, Nonlinear Dynamics, and the Heisenberg Limit. *Phys. Rev. Lett.* 102 100401 (2009).

[28] B. M. Escher, M. Filho, and L. Davidovich, General framework for estimating the ultimate precision limit in noisy quantum-enhanced metrology, *Nat. Phys.* 7, 406 (2011).

[29] N. Spagnalo, L. Aparo, C. Vitelli, A. Crespi, R. Ramponi, R. Osellame, P. Mataloni and F. Sciarrino, Quantum interferometry with three-dimensional geometry, *Sci. Rep.* 2, 862 (2010).

[30] M. Kacprowicz, R. Demkowicz-Dobrzański, W. Wasilewski, K. Banaszek and I. A. Walmsley, Experimental quantum-enhanced estimation of a lossy phase shift, *Nat. Photon.* 4, 357 (2010).

[31] G. Toth, Multipartite entanglement and high precision metrology, *Phys. Rev. A* 85, 022322 (2012).

[32] Y. Matsuzaki, S. C. Benjamin and J. Fitzsimons, Magnetic field sensing beyond the standard quantum limit under the effect of decoherence, *Phys. Rev. A* 84, 012103 (2011).

[33] S. Alipour, M. Mehboudi, and A.T. Rezakhani, Quantum Metrology in Open Systems: Dissipative Cramér-Rao Bound, *Phys. Rev. L ett.* 112, 120405 (2014).

[34] J. Liu, X. Jing and X. Wang. Phase-matching condition for enhancement of phase sensitivity in quantum metrology, *Phys. Rev. A* 88, 042316 (2013).

[35] R. Jozsa, D. S. Abrams, J. P. Dowling, and C. P. Williams, Quantum Clock Synchronization Based on Shared Prior Entanglement, *Phys. Rev. Lett.* 85, 2010 (2000).

[36] J. J. Bollinger, W. M. Itano, D. J. Wineland, and D. J. Heinzen, Optimal frequency measurements with maximally correlated states, *Phys. Rev. A* 54, R 4649 (1996).

[37] M.Tsang, Quantum metrology with open dynamical systems, New J. Phys. 15, 073005 (2013).

[38] M.Tsang, Ziv-Zakai Error Bounds for Quantum Parameter Estimation, Phys. Rev. Lett. 108, 230401 (2012).

[39] J. Liu, X. Jing, W.Zhong and X. Wang, Quantum Fisher Information for Density Matrices with Arbitrary Ranks, *Commun. Theor. Phys.* 61, 45-50 (2014).

[40] F. Ozaydin, A. A. Altintas, S. Bugu and C. Yesilyurt, Quantum Fisher Information of N Particles in the Superposition of W and GHZ States, *Int. J. Theor. Phys*, 52, 2977 (2013).

[41] F. Ozaydin, A. A. Altintas, S. Bugu, C. Yesilyurt, Behavior of Quantum Fisher Information of Bell Pairs under Decoherence Channels, *Acta Physica Polonica A*, 125 (2), 606 (2014).

[42] F. Ozaydin, A. A. Altintas, S. Bugu, C. Yesilyurt and M.Arik, Quantum Fisher Information of Several Qubits in the Superposition of A GHZ and two W States with Arbitrary Relative Phase International Journal of Theoretical Physics, DOI 10.1007/s10773-014-2119-4 (2014)

[43] P. Hyllus, O. Gühne, and A. Smerzi, Not all pure entangled states are useful for sub-shot-noise interferometry, *Phys. Rev. A* 82, 012337 (2010).

[44] L. Czekaj, A. Przysiezna, M. Horodecki and P. Horodecki, Quantum metrology: Heisenberg limit with bound entanglement, arxiv:1403.5867

[45] J. Ma, Y-X. Huang, X. Wang and C. P. Sun, Quantum Fisher information of the Greenberger-Horne-Zeilinger state in decoherence channels, *Phys. Rev. A* 84, 022302 (2011).

[46] J. Rehacek and Z. Hradil, Quantification of Entanglement by Means of Convergent Iterations, *Phys. Rev. Lett.* 90 127904 (2003).

[47] J. A. Miszczak, Generating and using truly random quantum states in Mathematica, *Comp. Phys. Comm.* 183, 1, 118 (2012).

[48] B. Cakmak and Z. Gedik, Quantum discord of SU(2) invariant states, *J. Phys. A: Math. Theor.* 46, 465302 (2013).

[49] A. I. Zenchuk, Unitary invariant discord as a measure of bipartite quantum correlations in an N-qubit quantum system, *Quant. Inf. Proc.,* 11, 1551 (2012).

[50] A. Peres, Separability Criterion for Density Matrices, *Phys. Rev. Lett.* 77, 1413 (1996).

[51] M. Horodecki, P. Horodecki, and R. Horodecki, Separability of Mixed States: Necessary and Sufficient Conditions, *Phys. Lett. A* 223, p. 1 (1996).

[52] F. Ozaydin, Phase damping destroys quantum Fisher information of W states, Phys. Lett. A 378, 3161 (2014).

| Class | Comparison with Maximized QFI |
|---|---|
| Entanglement($\rho_1$)=Entanglement($\rho_2$)=0 | MQFI($\rho_1$) > MQFI($\rho_2$) |
| | MQFI($\rho_1$) = MQFI($\rho_2$) |
| | MQFI($\rho_1$) < MQFI($\rho_2$) |
| Entanglement($\rho_2$)>Entanglement($\rho_1$)>0 | MQFI($\rho_1$) > MQFI($\rho_2$) |
| | MQFI($\rho_1$) = MQFI($\rho_2$) |
| | MQFI($\rho_1$) < MQFI($\rho_2$) |
| Entanglement($\rho_1$)=Entanglement($\rho_2$)>0 | MQFI($\rho_1$) > MQFI($\rho_2$) |
| | MQFI($\rho_1$) = MQFI($\rho_2$) |
| | MQFI($\rho_1$) < MQFI($\rho_2$) |
| Entanglement($\rho_1$)>Entanglement($\rho_2$)>0 | MQFI($\rho_1$) > MQFI($\rho_2$) |
| | MQFI($\rho_1$) = MQFI($\rho_2$) |
| | MQFI($\rho_1$) < MQFI($\rho_2$) |

Table 1. Ordering the general two-qubit states with respect to maximized quantum Fisher information and standard entanglement measures.

**Acknowledgments**

This publication has been funded by Isik University Scientific Research Funding Agency under Grant Number: BAP-14A101 and by Provus A MasterCard Company R&D Center. Authors would like to thank to Jaroslaw Miszczak for fruitful discussions.

**Author contribution statement**

V.E., F.O. and A.A.A. devised the model, designed the study. V.E. carried out the numerical simulations. V.E., F.O. and A.A.A. analyzed the data, prepared the figures and wrote the main text of the manuscript.

**Additional information**

**Competing financial interests:** The authors declare no competing financial interests.

**How to cite this article:** Erol, V., Ozaydin, F. & Altintas, A.A. Analysis of Entanglement Measures and LOCC Maximized Quantum Fisher Information of General Two Qubit Systems. Sci. Rep. 4, 5422; DOI:10.1038/srep05422 (2014).